\documentclass[manuscript,screen]{acmart}
\AtBeginDocument{%
  \providecommand\BibTeX{{%
    \normalfont B\kern-0.5em{\scshape i\kern-0.25em b}\kern-0.8em\TeX}}}

\copyrightyear{2024}
\acmYear{2024}
\setcopyright{rightsretained}
\acmConference[MOBILEHCI Adjunct '24]{26th International Conference on Mobile Human-Computer Interaction}{September 30-October 3, 2024}{Melbourne, VIC, Australia}
\acmBooktitle{26th International Conference on Mobile Human-Computer Interaction (MOBILEHCI Adjunct '24), September 30-October 3, 2024, Melbourne, VIC, Australia}
\acmDOI{10.1145/3640471.3680244}
\acmISBN{979-8-4007-0506-9/24/09}

\begin{document}

\title[An Initial Design Space of Domain-specific LVMs in HRI]{Vision Beyond Boundaries: An Initial Design Space of Domain-specific Large Vision Models in Human-robot Interaction}

\author{Yuchong Zhang}
\email{yuchongz@kth.se}
\affiliation{%
  \institution{KTH Royal Institute of Technology}
  \city{Stockholm}
  \country{Sweden}
}

\author{Yong Ma}
\affiliation{%
  \institution{University of Bergen}
  \city{Bergen}
  \country{Norway}}
\email{Yong.Ma@uib.no}

\author{Danica Kragic}
\affiliation{%
  \institution{KTH Royal Institute of Technology }
  \city{Stockholm}
  \country{Sweden}
\email{dani@kth.se}
}

\renewcommand{\shortauthors}{Zhang et al.}

\begin{abstract}
The emergence of large vision models (LVMs) is following in the footsteps of the recent prosperity of Large Language Models (LLMs) in following years. However, there's a noticeable gap in structured research applying LVMs to human-robot interaction (HRI), despite extensive evidence supporting the efficacy of vision models in enhancing interactions between humans and robots. Recognizing the vast and anticipated potential, we introduce an initial design space that incorporates domain-specific LVMs, chosen for their superior performance over normal models. We delve into three primary dimensions: HRI contexts, vision-based tasks, and specific domains. The empirical evaluation was implemented among 15 experts across five evaluated metrics, showcasing the primary efficacy in relevant decision-making scenarios. We explore the process of ideation and potential application scenarios, envisioning this design space as a foundational guideline for future HRI system design, emphasizing accurate domain alignment and model selection.
\end{abstract}





\maketitle



\section{Introduction}

The rapid advancement of robotics and human-computer interaction (HCI) has significantly progressed human–robot interaction (HRI), impacting daily life. Key research focuses on understanding interactivity and social behavior between robots and humans \cite{dautenhahn2007socially,zhang2024mind}, aiming to enable robots to predict human intent and efficiently complete tasks. Previous literature \cite{murphy2010human,kosuge2004human,sheridan2016human,zhang2024will} highlighted HRI's core aspects: human supervisory control of robots, cooperative task execution, and achieving social goals through interaction. Visual data is crucial in designing intuitive HRI contexts, including gestural interaction \cite{fan2022continuous}, object segmentation \cite{kim2010line}, and video tracking \cite{menezes2003visual}. Over the past decade, computer vision has evolved remarkably alongside advancements in deep learning. Deep neural networks have enabled significant breakthroughs in tasks like object detection, image segmentation \cite{rao2022monitoring}, and scene reconstruction \cite{wang2023review}. More recently, computer vision has seen progress in visual tracking, video captioning, pose estimation, and innovative content generation driven by the emergence of generative artificial intelligence (GenAI), with impressive applications across various domains. Large language models (LLMs) such as ChatGPT, have excelled in text-based tasks, inspiring the development of large vision models (LVMs) which are now pivotal in advancing vision-based analysis and interpretation \cite{rawte2023survey}.

The emergence of LVMs has revolutionized HRI by addressing long-standing challenges in visual perception and interpretation \cite{wang2023review}. Originally developed for image-based tasks like object recognition and scene understanding, LVMs are now applied in robotics, enabling the interpretation of complex scenes and informed decision-making with multimodal input. These models enhance robustness and efficiency, similar to how LLMs have transformed text-based applications \cite{zhang2023large}. However, unlike LLMs, which perform well across varied scenarios, LVMs require domain-specific training due to discrepancies in visual content between formal and informal platforms. Customized LVMs excel in comprehending the unique and nuanced visual content relevant to particular contexts in specific HRI systems. The landscape of AI, particularly in computer vision, has transformed significantly due to the advent of vision transformers and LVMs. Pioneering models like OpenAI's CLIP \cite{radford2021learning}, Landing AI's LandingLens \footnote{https://landing.ai/platform/}, Google's ViT \cite{dosovitskiy2020image}, and the SWIN Transformer \cite{liu2021swin} have revolutionized visual data processing. Domain-specific LVMs, such as OpenAI's GPT-4V \footnote{https://openai.com/research/gpt-4v-system-card} and Meta's SAM \cite{kirillov2023segment}, focus on specialized enterprise applications, enabling businesses to customize models for their specific needs. These domain-specific LVMs offer advantages over general models, including reduced training costs, enhanced accuracy, and unprecedented scalability through extensive parameter fine-tuning \footnote{https://www.solulab.com/large-vision-models/\#What\_are\_Large\_Vision\_Models}.

According to Landing AI\footnote{https://landing.ai/}, the LVM revolution lags behind LLMs by two to three years, but domain-specific LVMs outperform general ones. We anticipate regular advancements from these models, leading to more intelligent HRI systems and seamless robot integration into human society. To our knowledge, this work is the first to explore LVMs' potential in human–robot interaction, contributing an initial design space and empirical evaluation to guide the use of domain-specific LVMs in future HRI systems.

\section{Related Work}

\subsection{Design Space with HRI}

In HCI, a design space is a tool that highlights diverse possibilities for crafting specific artifacts, enriching and facilitating the design process. It evolves iteratively, incorporating new potentials based on insights from each design iteration \cite{maclean1991design}. Numerous studies have explored design spaces in HRI. In 2002, Dautenhahn investigated the design space of social robots, identifying niche spaces for human trust \cite{dautenhahn2002design}. Woods et al. developed a design space to identify children's perspectives on robots, highlighting 'Behavioral Intention' and 'Emotional Expression' as key factors \cite{woods2006exploring}. Walters proposed an empirical framework examining social robots' appearance and behavior, finding that consistent behavior and appearance are preferred by humans \cite{walters2008design}. Kalegina et al. conducted a survey on human perception of robot faces, noting that certain facial features significantly impact perceptions of a robot's capabilities and trustworthiness, influencing its appropriateness for specific tasks or environments \cite{kalegina2018characterizing}.

\subsection{LVMs in HRI: status}

A considerable amount of research has integrated advanced vision models into HRI systems for specialized tasks \cite{kang2023video,ong2012sensor,sheron2021projection,dehghan2019online,sakurai2007recognizing,martinez2001clustering,xu2020multi,lee2016hardware,zhao2018collaborative,van2024puppeteer}. Object detection is a prevalent task, with vision models enhancing performance and enabling better robot-human interaction \cite{hsu2018human,shieh2013fast,latif2023human,maiettini2021weakly,saputra2019real}. Azagra et al. proposed a pipeline for interaction type recognition and target object detection \cite{azagra2017multimodal}. Fang et al. developed a hand-held object detection framework using YOLO \cite{redmon2016you} and KCF \cite{henriques2014high} on RGB-D video data \cite{fang2018understanding}. Yu et al. used the LSTM model \cite{hochreiter1997long} for video object detection in a space robot-human scenario with gestural interactions \cite{yu2022deep}. Visual segmentation, which partitions images and videos, is another common task in HRI research \cite{chavarria2013simultaneous,ghidary2001multi,fei2023hybrid}. \"{U}ckermann et al. presented a real-time 3D segmentation algorithm for robot grasping with a Shadow Robot Hand \cite{uckermann2013realtime}. Fan et al. proposed a gesture segmentation and recognition architecture using deep convolutional neural networks \cite{fan2022continuous}. Kim et al. designed a multimodal HRI framework for online object segmentation and gesture recognition \cite{kim2010line}. Besides, visual tracking is also frequently used in HRI research. My et al. \cite{my2013real} and Putro et al. \cite{putro2018real} developed real-time face tracking systems for seamless interaction with moving robots, while Song et al. created a face tracking method using image data in a human-mobile robot environment \cite{song2004face}. Other tasks like pose estimation, content generation, and visual recognition are also common in HRI studies. However, there is a notable lack of research integrating well-defined LVMs into human-robot interaction, especially domain-specific LVMs. Given the effectiveness of LLMs in text-based information processing, we posit that domain-specific LVMs could transform vision-based contextual analysis across various industries by leveraging their specialized expertise.


\begin{figure*}[!t]
  \centering
  \includegraphics[width=\linewidth]{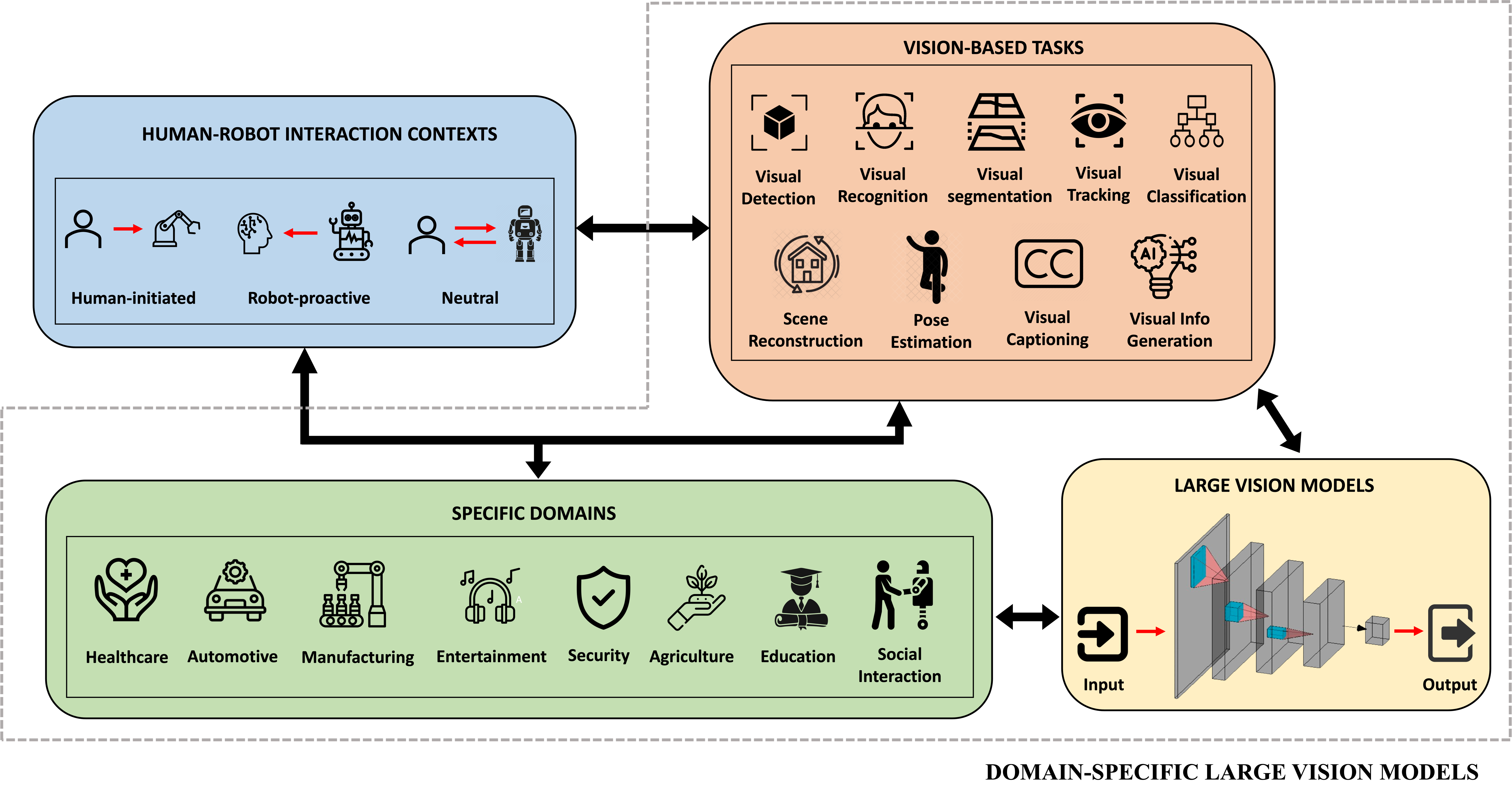}
  \caption{Our proposed design space. The HRI contexts interplay with vision-based tasks tailored to specific domains. Corresponding LVMs deliver performance that meets the unique requirements of each case.}
  \label{ds}
\end{figure*}

\section{The Design Space}
In this section, we present the formulation of the proposed design space encapsulating three main dimensions.

\subsection{HRI Contexts}

HRI has become a crucial field at the intersection of robotics and HCI over the past decades. As robots increasingly coexist with humans in various work environments, creating safe and user-friendly systems is essential. The challenge lies in integrating human behaviors into robots' decision-making frameworks for close physical interactions. With AI technology advancing, HRI research focuses on ensuring safe physical interactions and fostering socially appropriate interactions sensitive to cultural nuances. The goal is to develop robots capable of intuitive communication with humans using speech, gestures, and facial expressions. HRI extends to multiple domains, including industrial automation, medical assistance, and personal companionship. Various studies have categorized different contexts of interactions between humans and robots, yielding similar outcomes based on distinct underlying principles \cite{schulz2018preferred,butepage2017human,sheridan2016human,xia2024shaping}. In our paper, we propose a new and concise categorization of HRI contexts tailored for future LVM deployment (see Figure ~\ref{ds}).

\begin{itemize}
    \item \small \textbf{\textit{Human-initiated:}} In this context, human supervisory control over robots is initiated for executing routine tasks. It is anticipated that humans will make decisions about the manner and timing of actions, subsequently directing, instructing, or commanding the robots to carry out these actions and provide relevant feedback. Under this framework, robots are considered passive entities, springing into action only upon receiving explicit control signals from human operators.
    \item \textbf{\textit{Robot-proactive:}} In this context, robots proactively determine the appropriate ways and timing of actions, and also guide humans on their following behaviours. The robots autonomously identify the necessary actions to progress to the subsequent steps required for routine tasks. Concurrently, they independently monitor the situation and observe human actions. Should the predefined goal not be achieved within a set time frame, the robots will then commit an action to ensure the accomplishment of the objective.
    \item \textbf{\textit{Neutral:}} In this context, both humans and robots proactively initiate actions, collaboratively deciding on the how and when of task execution to achieve mutual goals. Actions required for progressing to the next sub-task are determined simultaneously by both parties. Robots select actions as they would in a robot-proactive context, but they also inform humans about their anticipated contributions, based on the robot's task plan. A key application of this approach is in social interactions between humans and robots, where robotic devices are designed to entertain, educate, comfort, and assist specific human users.
\end{itemize}

\subsection{Vision-based tasks}
The second dimension identified pertains to vision-based tasks, developed by synthesizing insights from published works on visual engagement in HRI and current categorizations of computer vision tasks \cite{voulodimos2018deep} (Figure ~\ref{ds}).

\begin{itemize}
    \item \small \textbf{\textit{Visual Detection:}} At present, this is the most prevalent task designed in contemporary HRI research employing vision models, such as object detection \cite{hsu2018human,shieh2013fast} and face detection \cite{lee2016hardware,putro2018real}. In the majority of cases, the visual detection tasks have significantly enhanced the robustness of interaction with robots.
    \item \textbf{\textit{Visual Recognition:}} This task is widely-adopted particularly in gesture-based interactive frameworks \cite{gandarias2018enhancing,chavarria2013simultaneous}, where the gestural recognition ensures the smooth control of the robots.
    \item \textbf{\textit{Visual Segmentation:}} Similarly, segmenting a specific set of areas of interest \cite{zhang2020automated} is currently another important focus used for instructing the robots \cite{uckermann2013realtime,kim2010line}.
    \item \textbf{\textit{Visual Tracking:}} In this task, human motion tracking, human behaviour tracking, and human body tracking are the most used instances for interacting mainly with mobile robots and humanoid robots \cite{menezes2003visual,molina2005real}.
    \item \textbf{\textit{Visual Classification:}} As a fundamental task that has evolved alongside deep learning, this area is anticipated to bring significant advantages to HRI systems, particularly in instances like human/robot intention classification \cite{ong2012sensor} and object classification in robot teaching \cite{dehghan2019online}.
    \item \textbf{\textit{Scene Reconstruction:}} The sophisticated task of reconstruction has seen rapid development in recent years. Several studies have already leveraged it for decision-making and proactive collaboration within the human-robot loop \cite{zhang2023part,fan2022vision}.
    \item \textbf{\textit{Pose Estimation:}} This task is often integrated with visual tracking, enabling robots to proficiently perform 3D estimation of either humans or objects \cite{xu2020multi}.
    \item \textbf{\textit{Visual Captioning:}} As an emerging vision task, it is expected to bring more convenience where robots provide education, or social companionship \cite{kang2023video}.
    \item \textbf{\textit{Visual Info Generation:}} With the swift advancements in GenAI, this task has demonstrated its high capability in the field of robotics in recent years. In HRI, we foresee its more utilization, building on the current successes in areas like motion generation \cite{zhao2018collaborative} and location generation \cite{saputra2019real}.
\end{itemize}

\subsection{Specific domains}

Compared to generic LVMs, domain-specific models require only 10\% to 30\% of the labeled data and produce significantly fewer errors. This advancement reduces computational demands and improves performance. We identified eight domains (Figure ~\ref{ds}) expected to encompass most current and prospective HRI application areas, based on insights from published works and research trends.

\begin{itemize}
    \item \small \textbf{\textit{Healthcare:}} HRI is revolutionizing health such as robotic-assisted surgery, where precision and reduced invasiveness are critical \cite{lee2012technical,esterwood2020personality}. Robots like exoskeletons are also being deployed for rehabilitation, providing physical therapy with adaptive routines. The superiority of LVMs in detecting anomalies in medication-related imaging and forecasting disease progression will enhancing the decision-making.
    \item \textbf{\textit{Automotive:}} LVMs are central in advancing the autonomous vehicles \cite{heydaryan2018safety,basu2016trust}. The real-time object and pedestrian detection, will enable safer navigation and interaction with the environment. Furthermore, LVMs enhance the design of ergonomic vehicle interiors, thereby elevating user experience and safety in driver-assist systems.
    \item \textbf{\textit{Manufacturing:}} LVMs enable robots to execute precise quality inspections, detect defects, and ensure accurate assembly, collaborating with human operators \cite{marvel2020towards,sadrfaridpour2017collaborative,tan2009human}. In addition, the process monitoring and maintenance significantly improve the efficiency and workplace safety.
    \item \textbf{\textit{Entertainment:}} LVMs in HRI are anticipated to vastly improve entertainment experiences through their prowess in generating visual content, including enhanced video and gaming experiences \cite{budiharto2017edurobot,pollmann2023entertainment}. Moreover, the capacity for analyzing visual data, like predicting human intentions, is set to further enrich the quality of interactive experiences between humans and robots.
    \item \textbf{\textit{Security:}} In critical situations, robots are enabled to perform continuous surveillance, recognize suspicious activities, and identify potential threats with desirable accuracy \cite{agrawal2017robot,lopez2017robotman}. Also, utilizing advanced facial recognition and behavior analysis from LVMs to ensure human safety and security publicly.
    \item \textbf{\textit{Agriculture:}} LVMs enable agricultural robots (such as drones) to accurately identify and classify crops, assess plant health, predict incoming yields, and detect pests or diseases, optimizing the cultivating and harvesting \cite{vasconez2019human,adamides2017hri}.
    \item \textbf{\textit{Education:}} This domain has been ubiquitously probed in HRI. With the aid of LVMs, educational robots are to precisely recognize and respond to manifold gestures and expressions from humans, facilitating more engaging and personalized teaching performance \cite{tekerek2009human,cui2022human}. Some vision-related techniques such as virtual reality/augmented reality \cite{zhang2023virtuality,zhang2021supporting,nowak2021augmented} are able to make learning more immersive and accessible.
    \item \textbf{\textit{Social Interaction:}} Social robotics is a crucial facet of HRI \cite{yan2014survey,kanda2017human,lemaignan2017artificial}. In social settings, LVMs enable robots to interpret and respond to human emotions and social cues, fostering more natural and empathetic communication and engagement. A wide-used case is eldercare, where the social robots can provide companionship and emotional support, tailored to individual needs.
\end{itemize}



\section{Evaluation}

\subsection{Study Design}

\begin{figure*}[!t]
  \centering
  \includegraphics[width=\linewidth]{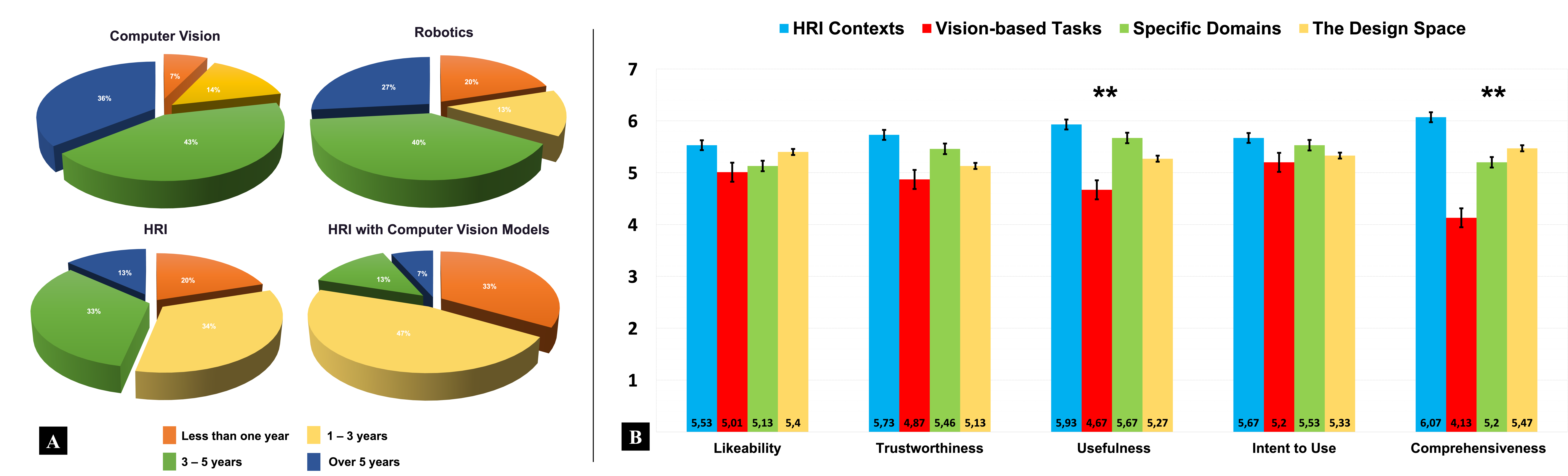}
  \caption{A: Demographics: academic background information of all participants. B: The outcomes of the five evaluated metrics, which assessed the three dimensions and the overall design space. **: $p<0.05$.}
  \label{results}
\end{figure*}

Exploring the design space's dimensions provides valuable insights into LVM diversity for HRI researchers and practitioners. However, it doesn't reveal how these variations affect individual responses. To address this, we conducted an expert evaluation to gather empirical evidence for developing HRI systems that evoke specific reactions to LVMs. We engaged 15 participants aged 24 to 42 (7 male, 6 female, 2 prefer not to say, M = 30.33, SD = 4.66) through email invitations and personal outreach. Participants had several years of experience in relevant areas (see Figure ~\ref{results}.A), enhancing the study's reliability. After a brief introduction to the design space, we used a questionnaire to explore experts' perspectives on its three dimensions and the design space itself. This study employed a questionnaire-based evaluation and quantitative analysis to measure five perceived metrics from existing literature \cite{gries2004methods,kalegina2018characterizing,baartmans2020social,haggerty2011comprehensiveness,zhang2023playing,zhang2023see,zhang2021affective}. The questionnaire included demographic information, an overview of the design space (see Figure ~\ref{ds}), and targeted questions to gauge perceptions of likeability, trustworthiness, usefulness, intent to use, and comprehensiveness, using a 7-point Likert scale. The questions are detailed in the supplementary material.

\subsection{Data Analysis}

On average, participants completed each trial in about 15 minutes, including a brief interview session to share their opinions. All participants successfully finished the study and received a small gift as appreciation. Before conducting statistical analysis, normality tests confirmed that all metrics adhered to a normal distribution ($p<0.001$). We used one-way ANOVA with repeated measures ($p<0.05$) for the five metrics, wherein, each dimension and the design space received satisfactory ratings across all metrics (Figure ~\ref{results}.B). Notably, the HRI context dimension achieved the highest ratings in every metric, while the vision-based tasks dimension received the lowest.

\renewcommand{\labelitemi}{--}
\begin{itemize}
    \item \small \textit{Likeability.} Participants showed an overall preference favoring the HRI contexts dimension as the most likable, closely followed by the design space itself and then the specific domains dimension. However, the differences in likeability ratings did not reach statistical significance.
    \item \textit{Trustworthiness.} In terms of trustworthiness, perceptions varied slightly from likeability. Here, the specific domains dimension was considered second only to the HRI contexts dimension, though the gap was marginal. Again, these differences were not statistically significant.
    \item \textit{Usefulness.} The vision-based tasks dimension was seen as the least motivational for future research, distinctly lagging behind the other dimensions. This assessment was supported by a significant statistical difference, as indicated by a repeated measures ANOVA ($(F(3, 42) = 4.145, p < 0.05)$), highlighting the variance in perceived usefulness across dimensions.
    \item \textit{Intent to Use.} The variance in the intention for potential adoption was the smallest across all dimensions with the design space itself when compared to other five evaluated metrics. The vision-based tasks dimension was rated only slightly falling behind other dimensions in this metric, although no significance was detected.
    \item \textit{Comprehensiveness.} Within the whole study, the HRI contexts dimension received the highest rating for comprehensiveness, nonetheless, the vision-based tasks dimension obtained the lowest rating. Statistical significance was affirmed by the repeated measures ANOVA on the comprehensiveness scales with $(F(3, 42) = 5.119, p < 0.05)$.
\end{itemize}


\subsection{Summary of Qualitative Feedback}
The majority of participants expressed positivity of the design space and its three dimensions. One participant, whose research interests align closely with the proposed topic, mentioned, \textit{"The design space you've introduced seems like it will be a useful tool and guide for my future research endeavors to a significant extent, as it clearly lists the essential components."} In terms of the three dimensions, another participant appreciated the strategy of outlining specific domains of focus before diving into practical research challenges brought by LVMs: \textit{"This approach of summarizing key domains before tackling practical research problems can streamline the process of identifying the precise area of interest for applying the vision model. It not only saves time but also leads to enhanced outcomes."} Moreover, a participant with years of experience in computer vision and robotics commented on the utility of the HRI contextual categorization, stating, \textit{"This categorization of contexts will aid in laying down the essential structure for the interactive system I plan to develop in the future, especially since my prior experience in this area is limited."}

\section{Discussion}
In this section, we delve into various facets of the proposed design space.

\textbf{It serves as a preliminary tool for designers and researchers.} In this study, we formulated an initial design space, featuring three dimensions tailored for users to develop future HRI systems using domain-specific LVMs. Our evaluation process included an empirical testing by experts and the demonstration of its foundational utility across six distinct metrics. Notably, the dimension focusing on HRI contexts received the highest appreciation, whereas the vision-based tasks dimension was evaluated as the least effective according to our metrics. This discrepancy might stem from experts' perception that, despite our efforts to be exhaustive, the list of vision tasks might not fully encapsulate the breadth of existing research in this area. Given the fast-paced advancements in computer vision, it's plausible that more innovative and effective vision tasks will emerge. Nevertheless, we believe that the interaction contexts between humans and robots can be effectively categorized within the three classes we proposed, based on the current state of knowledge. The overall design space received satisfactory appraisal in all metrics, implying that its fundamental utility was primarily endorsed by experts. 

\textbf{The ideation is based on the existing work and emerging trends.} While numerous studies have focused on integrating various vision models in HRI, primarily for enhanced visual detection and robot control, there is a lack of structured approaches in utilizing LVMs for efficient interaction between humans and robots. Given the current surge in LLMs, we anticipate a similar revolutionary trajectory for LVMs in following years. The concept of domain-specific LVMs, first promoted and advocated by Landing AI, has shown superior performance over traditional models. This underscores the need for a foundational guideline that assists in designing future HRI systems, delineating specific contexts, vision tasks, and domains.

\textbf{The proposed approach offers several advantages over current models.} Utilizing domain-specific unlabeled data substantially reduces the reliance on expensive labeled data, making the development of models more economical. This cost reduction could lead to enhanced interactions with robots. Additionally, the need for less data is capable to yield faster training periods, enabling the rapid execution and implementation of LVMs in real-world HRI applications especially in wearable and social robotics. Furthermore, by selecting datasets that are specialized for specific domains, the models are expected to achieve significantly higher accuracy due to a deeper understanding of the intricacies unique to different areas.

\textbf{Using the design space.} Our proposed design space serves as an effective tool for preliminarily identifying key elements in human-robot collaborative environments. For instance, the development of a robot arm controlled by human gestures to grasp objects includes: a human-initiated context is established, focusing on tasks such as gesture recognition (visual recognition) and object detection (visual detection). The selection and application of LVMs then follow, tailored to the domains (such as education or entertainment), enhancing visual task performance and facilitating smoother control and interaction. Similarly, in a social companion robot scenario, interaction occurs within a neutral context, employing LVMs specialized for social interaction domain. These models can be utilized for recognizing human emotions and behaviors (visual recognition) and providing empathetic responses (visual captioning). Through our expert evaluation, we are expecting that our design space will aid in informed decision-making in relevant cases.

\textbf{General challenges.} Despite their considerable promise, several challenges need addressing for the broader adoption and practical application. A primary issue is accurately determining the appropriate domain before utilizing specific LVMs to guarantee the precision of visual data, as many HRI studies often overlook domain specification. Data availability may be constrained in some fields due to the difficulty of accessing substantial amounts of domain-specific data, particularly in certain industries. Furthermore, establishing effective interaction methods with robots remains an ongoing challenge in the broader field of HRI research.

\textbf{Ethical considerations.} Another significant factor impacting the successful deployment of the design space is ethic concerns. LVMs require a vast number of database, which can raise a significant factor -- data bias and fairness, due to the bias inheriting in the training data. This can lead to inequitable or unethical outcomes, especially in sensitive areas like facial recognition in security and healthcare, where privacy concerns are paramount. Additionally, the interpretability and explainability of these large models are crucial; a lack of understanding of how these models function could compromise transparency. The substantial computational resources required for LVMs pose another significant gap, potentially impeding the successful deployment of an effective HRI system. Finally, ethical issues from HRI's perspective, such as robots being safe and trustworthy rather than threatening data privacy and human well-being, are expected to be addressed \cite{etemad2022ethical}.

\textbf{Limitations of the initial design space.} While our design space is formulated on existing HRI research and the evolving trends of large foundational models, certain limitations are inherent. We have identified nine vision-based tasks and eight domains aimed at encompassing the most relevant scenarios. However, the continuous advancement of vision models may give rise to new visual tasks in the future, and HRI could potentially expand into currently not-well-explored domains such as pharmacy and electronics. Moreover, the absence of user studies or expert evaluations in our approach means a lack of practical feedback and insights from professionals in the field, which is crucial for comprehensive evaluation and enhancement of the design space.

\normalsize
\section{Conclusion}
In this paper, we present an initial design space focused on the future development of HRI systems, leveraging domain-specific LVMs. Mirroring the recent success of LLMs, we foresee LVMs as transformative agents in vision-based information processing within HRI endeavors. We carried out an empirical evaluation with 15 expert participants, and all perceived metrics pointed towards the promising employment of our proposed methodology. Our advocacy for domain-specific models instead of the normal ones stems from their potential to drive significant enhancements and progress in targeted scenarios, enabling more fluid interactions with robots. We believe that this design space will serve as an accelerator for innovative HRI designs across a variety of domains.

\begin{acks}
This work was supported by the Swedish Foundation for Strategic Research (SSF) grant FUS21-0067.
\end{acks}

\bibliographystyle{ACM-Reference-Format}
\bibliography{sample-base}






\end{document}